\DeclareUnicodeCharacter{0301}{\'{e}}
\documentclass[superscriptaddress,reprint,amsmath,amssymb,showkeys]{revtex4-2}
\setcitestyle{super}
\usepackage{graphicx}
\usepackage{dcolumn}
\usepackage{bm}
\usepackage{color}
\usepackage{float}
\usepackage[utf8]{inputenc}
\usepackage[mathlines]{lineno}
\usepackage{multirow}
\usepackage{amsmath}
\usepackage{xr-hyper}
\usepackage{hyperref}
\usepackage{soul,xcolor}
\usepackage{makecell}
\usepackage{amssymb }

\soulregister\cite7
\soulregister\ref7
\setstcolor{blue}
\usepackage[version=3]{mhchem} 
\usepackage{xcolor}
\usepackage{float}
\usepackage{parskip}
\usepackage{svg}

\begin{document}

\title{Tapered optical fibers coated with Rare-Earth complexes for quantum applications}

\author{Ori Ezrah Mor}
\affiliation{Department of Chemical and Biological Physics, Weizmann Institute of Science, Rehovot 7610001, Israel}
\author{Tal Ohana}
\affiliation{Department of Chemical and Biological Physics, Weizmann Institute of Science, Rehovot 7610001, Israel}
\author{Adrien Borne}
\affiliation{Department of Chemical and Biological Physics, Weizmann Institute of Science, Rehovot 7610001, Israel}
\affiliation{Current affiliation: Université de Paris, Laboratoire Matériaux et Phénomènes Quantiques, CNRS UMR7162, F-75013 Paris, France}
\author{Yael Diskin Posner}
\affiliation{Department of Chemical Research Support, Weizmann Institute of Sceince, Rehovot, Israel}
\author{Maor Asher}
\affiliation{Department of Chemical and Biological Physics, Weizmann Institute of Science, Rehovot 7610001, Israel}
\author{Omer Yaffe}
\affiliation{Department of Chemical and Biological Physics, Weizmann Institute of Science, Rehovot 7610001, Israel}
\author{Abraham Shanzer}
\affiliation{Department of Molecular Chemistry and Material Science, Weizmann Institute of Science, Rehovot, Israel}
\author{Barak Dayan}
\affiliation{Department of Chemical and Biological Physics, Weizmann Institute of Science, Rehovot 7610001, Israel}

\begin{abstract}
Crystals and fibers doped with Rare Earth (RE) ions provide the basis to most of today’s solid-state optical systems, from lasers and telecom devices to emerging potential quantum applications such as quantum memories and optical to microwave conversion.
The two platforms, doped crystals and doped fibers, seem mutually exclusive, each having its own strengths and limitations – the former providing high homogeneity and coherence, and the latter offering the advantages of robust optical waveguides.
Here we present a hybrid platform that does not rely on doping but rather on coating the waveguide - a tapered silica optical fiber - with a monolayer of complexes, each containing a single RE ion.
The complexes offer an identical, tailored environment to each ion, thus minimizing inhomogeneity and allowing tuning of their properties to the desired application. 
Specifically, we use highly luminescent Yb$^{+3}[\mathrm{Zn(II)}_\mathrm{MC}$ (QXA)] complexes, which isolate the RE ion from the environment and suppress non-radiative decay channels. 
We demonstrate that the beneficial optical transitions of the Yb$^{+3}$ are retained after deposition on the tapered fiber, and observe an excited-state lifetime of over 0.9~ms, on par with state-of-the-art Yb doped inorganic crystals.
\end{abstract}

\date{This manuscript was compiled on \today}
\maketitle
\section*{Introduction}
Mapping quantum states onto the hyperfine states of rare-earth (RE) ions is one of the promising platforms for quantum technologies in general, and for interaction with photonic qubits in particular. 
The partially-filled 4f shell of all the RE$^{+3}$ ions is shielded from the environment by the outer 5s and 5p shells, thereby reducing the influence of the host lattice on intra-shell f-f transitions. 
As a result, RE ions exhibit sharp absorption and emission spectral lines at wavelengths that range from microwave to optical and UV, even when located inside a solid-state matrix or an organic complex. 
Accordingly, RE-doped crystals are the key building block in most solid-state lasers and amplifiers, and RE-doped fibers provide the basis for practically all fiber lasers\cite{koester1964amplification,fu2017review,jackson2012towards}.
\footnote{Note that small concentration of RE ions has been detected even in undoped fibers, as reported in \cite{wolfowicz2021parasitic}}
Notably, the two major platforms for interfacing RE ions with light rely on doping the host material - either crystals or fibers. Each of these platforms exhibits its own physical properties and advantages. 
While RE-doped crystals enable low inhomogeneity and high coherence, RE-doped fibers offer the efficiency and practicality of a confined optical waveguide.
Efforts to reconcile the two approaches have proved to be nontrivial. 
RE-doped silica fibers exhibit charge diffusion and tunneling modes that induce large inhomogeneity and reduce the coherence time, even at cryogenic temperatures ~\cite{broer1986low,jin2015telecom,geva1998optical}. 
Fabrication of waveguides based on RE-doped crystals requires methods such as focused ion beam milling, ion diffusion, laser writing\cite{saglamyurek2011broadband,thiel2012rare,sinclair2017properties,chen2007development,seri2018laser,zhong2015nanophotonic}, and epitaxial growth of RE doped silicon \cite{weiss2021erbium,gritsch2021narrow}. 
The effort of coherently incorporating RE ions into a single mode waveguide platform is perhaps mostly significant in the field of quantum information, where RE-doped crystals at cryogenic temperatures have already been harnessed to demonstrate quantum optical memories \cite{saglamyurek2011broadband,seri2018laser,thiel2012rare,afzelius2010demonstration,hedges2010efficient,longdell2005stopped}, and are also the basis for a number of proposals for optical to microwave qubit conversion \cite{bartholomew2019on}.
One of the most advanced platforms in this direction is a nano-photonic cavity fabricated in a YVO$_4$ crystal doped with Nd$^{+3}$ or Yb$^{+3}$ ions exhibiting long coherence time and narrow inhomogeneous broadening \cite{zhong2015nanophotonic}, which has been harnessed to demonstrate a quantum memory \cite{zhong2017nanophotonic}, a cavity protected interface between RE and photonic qubits \cite{zhong2017interfacing}, a readout of photonic qubit stored in a single ion in a single shot measurement \cite{kindem2020control}, and an optical interface between RE ions in a cavity and adjacent nuclear spins for quantum memory and entangled states \cite{ruskuc2022nuclear}.
Additionally, Purcell enhancement was demonstrated with a ring resonator fabricated in Yb-doped silicon nitride \cite{ding2016multidimensional}.
\begin{figure*}[ht]
  \centering
        \includegraphics[width=16 cm]{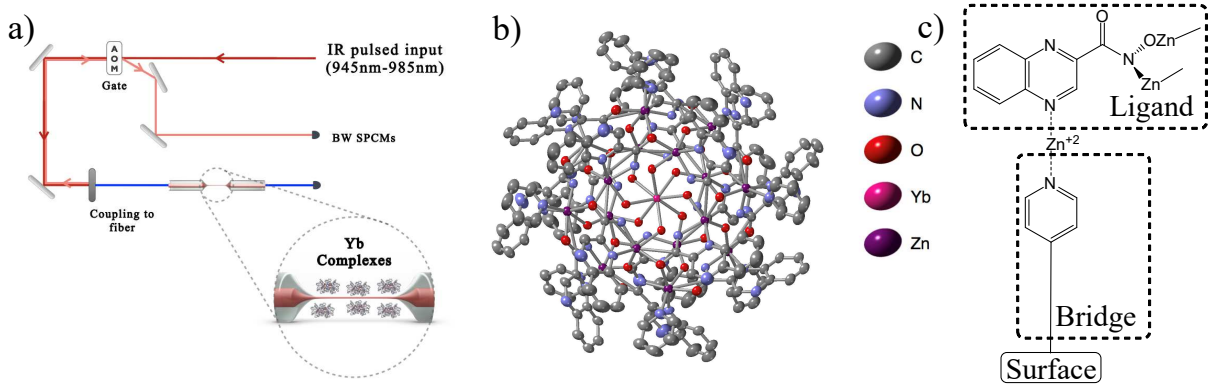}
  \caption{The device of this study: (a) a tapered single-mode fiber coated by a monolayer of tailored rare-earth (RE) complexes, each containing a single RE ion, is spliced to the optical setup for Photoluminescence Excitation (PLE) spectroscopy: $\sim 1 \mu s$ pulses are generated from a tunable laser (945-985 nm) and sent to the TOF (red arrow).
    The PL signal is collected on SPCMs in the backward (BW) direction (light red arrow) and is separated from parasitic laser reflections using a single pass AOM serving as a gate. (b) the structure of the complex, Yb$^{+3}[\mathrm{Zn(II)}_\mathrm{MC}$ (QXA)], obtained by XRD (thermal ellipsoids structure presentation with probability of $50\%$; 
  Hydrogens are omitted for clarity).
  (c) For binding the complex to the surface, the latter is first functionalized with a pyridyl end-group, then $Zn^{+2}$ salt is used to bridge the surface and the vacant pyridyl nitrogen of the complex (See Fig. 1 in the supporting information (SI) for a detailed description).}
  \label{device}
\end{figure*}

Recently, molecular complexes have been considered a promising host for metal ion qubits, such as transition metal spins \cite{bayliss2020optically,fataftah2020trigonal,wojnar2020nickel}, and also for RE ions \cite{kumar2021optical,serrano2021rare}.
They offer control over the immediate environment of the ions at the single-atom level, thus the optical properties can be tuned as necessary. 
In particular, the approach enables a uniform environment to all ions - a desirable property for quantum applications.
A key difference between RE complexes and RE doped crystals is the presence of high-frequency vibrations in the complexes' structure, which can quench the excited state and reduce the quantum yield, namely introduce dissipation and reduce the probability of an optical emission\cite{tan2006quenching}. 
Therefore, highly luminescent RE complexes are designed to minimize the number of high-frequency vibrations (mainly X-H bonds with frequency $ > 3000$~$cm^{-1}$) in the structure, and to maximize the distance between them to the RE ion. 

Here we present a new approach for coupling RE complexes with tapered optical fibers (TOF).
TOFs allow optical interface with matter in their vicinity through the evanescent field.
Such fibers have been under active research in the field of quantum applications and spectroscopy, particularly with atoms trapped near the surface \cite{meng2018near,sayrin2015storage} as well as crystals of molecular dyes addressing a single dye molecule\cite{skoff2018optical}. 
We coat the TOF with Yb$^{+3}[\mathrm{Zn(II)}_\mathrm{MC}$ (QXA)] complexes (Fig.~\ref{device}a and b) which emit at near infra-red (NIR).
We form a self assembled monolayer (SAM) of our complex by functionalizing the surface of the TOF with a pyridyl end-group (Fig.~\ref{device}c).
An important benefit of our approach is that it does not involve doping of the raw material or custom-fabrication process of the waveguide.\\

Since, in contrast to doped inorganic crystals, non-radiative loss is a major issue when discussing complexes, we consider the applicability of this device to quantum applications in terms of its intrinsic quantum yield. Specifically, we present optical measurements exhibiting PL decay times of over 0.95~ms, which are comparable to reported devices that are based on inorganic crystals.
\section*{Results and discussion}
Our work includes synthesis and crystallization of a specially designed Yb complex, and formation of a SAM on the surface of a TOF, resulting in record PL lifetimes in comparison to other reported Yb organic complexes. 
In the next subsection we discuss the synthesis and characterization, followed by the formation of the SAM, and finally we describe the luminescence properties of the bulk crystals and the functionalized TOF.

\subsection*{Synthesis and Characterization of Yb$^{+3}[{Zn(II)}_{MC}$(QXA)]}
The main quenching mechanism  of the luminescence originates from coupling to high-energy vibrations such as C-H, O-H, and N-H which are common in organic ligands~\cite{tan2006quenching}.
Therefore, for high intrinsic quantum yield, H atoms must not be adjacent to the ion. 
An additional property desired for quantum information is a relatively large separation between adjacent Yb ions in order to reduce dipole-dipole interactions between them.
Considering the above, we based our device on the Metallacrown family of complexes which was first synthesized by Pecoraro et al.~\cite{lah1989isolation}.
These complexes consist of rings formed by heavier atoms – a transition metal (such as Zn$^{+2}$), oxygen, and nitrogen- reminiscent to crown ethers.
As a result, hydrogen atoms are absent from the central ion vicinity, thus non-radiative decay channels are suppressed~\cite{trivedi2014highly}.
In order to allow binding to the surface of the TOF, we synthesized a modified version of the Metallacrown complex based on the 2-Quinaldichydroxamic acid (QHA)~\cite{trivedi2014highly}, where we used the 2-Quinoxalinehydroxamic acid (QXA) ligand~\cite{ivana2018selective,eliseeva2022tuning} that has an additional vacant nitrogen atom, which can be harnessed for surface binding.
We synthesized the ligand from its precursor 2-Quinoxalinecarboxylic acid, and the corresponding Metallacrown complex via a modified procedure from Ref.~\cite{trivedi2014highly}. 
A detailed description of the synthesis is given in section A of the supporting information (SI). 
The complexes were crystallized using vapor diffusion of ethyl acetate into a solution of the complexes in N,N-dimethylformamide (DMF) and pyridine~\cite{trivedi2014highly}.
\begin{figure}[h!]
    \centering
    \includegraphics[width=8 cm]{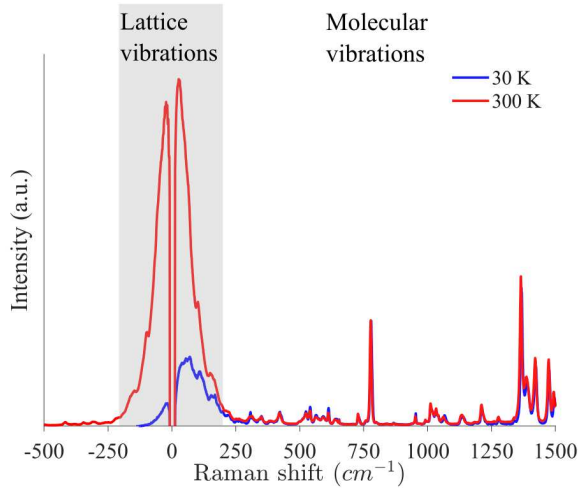}
    \caption{Temperature dependent Raman spectra of Yb$^{+3}[\mathrm{Zn(II)}_\mathrm{MC}$ (QXA)] crystalline bulk, under excitation at $\lambda_{ex}=785$ nm at room and at cryogenic temperatures (red and blue respectively).
    The spectrum is divided into two regimes: The low frequency (10-200 $cm^{-1}$) [gray], and the high frequency molecular vibrations ($> 200$ $ cm^{-1}$)[white].
    The spectra were normalized to the peak at $778 cm^{-1}$}
    \label{fig_raman}
\end{figure}
X-ray diffraction (XRD) (conducted on the crystals without drying; Provided in section B in the SI) indicates that the complexes form tetragonal crystals and exhibit an isostructure of their QHA analogue: two 4-membered rings bound to the central RE ion with a middle 8-membered ring.
The resolved structure of the complex is presented in Fig.~\ref{device}~b. 
Raman spectroscopy of the crystalline bulk (i.e. dried single crystals) (Fig.~\ref{fig_raman}), reveals sharp high frequency ($>$250~cm$^{-1}$) modes attributed to the organic ligand.
Importantly, the low frequency (10-200~cm$^{-1}$) Raman spectra exhibits a broad and diffused signal both at room and cryogenic temperatures (gray).  
Such a broad Raman signal in the low frequency range indicates that the crystalline bulk exhibits a relatively large level of disorder.
This is important because it indicates that the relatively sharp luminescence spectrum of the bulk (shown below in Fig.~\ref{ref_pl_ple}) is not related to the long range order but to the electronically isolated nature of the Yb ion. 
 \subsection*{Formation of Yb$^{+3}[\mathrm{Zn(II)}_\mathrm{MC}$(QXA)] SAM on~the~TOF}
As illustrated in Fig.~\ref{device}c, the interface between the silica and the Yb$^{+3}[\mathrm{Zn(II)}_\mathrm{MC}$ (QXA)] complex is generated utilizing a bi-functional layer that is silylated on one end, and possessing a pyridyl moiety on the other~\cite{de2014pyridine,altman2006controlling}.
Since the tapered optical fiber is frail, the functionalization procedure was performed at ambient conditions, to avoid mechanical damage and optical loss.
The cascaded procedure involves the formation of three layers: the template layer, the zinc layer, and the complex layer. 
The surface is first functionalized with a silane reagent containing a chain of 7 atoms and ends with a pyridyl ring (the template layer).
The pyridyl ring is then connected to a $Zn^{+2}$ salt layer, which allows the binding of the complexes via the vacant quinoxaline nitrogen atoms.
The detailed procedure is provided in section C of the SI.
To characterize the SAM, we first functionalized a Si substrate.   
After forming the SAM, we used an AFM tip operating in a contact mode to remove an area of 0.25~$\mu m^2$ followed by a tapping mode imaging of the removed area (Fig. 2 top in the SI).
This procedure allowed us to measure the SAM thickness.
As shown in Fig. 2 bottom in the SI, a typical layer is of approximately 3~nm thickness, which is compatible with a 1.5~nm sized complex that lies on a chain of 12 atoms.
It is important to note that the thickness of the SAM varied between 1.8 and 8~nm but this is a reasonable variation for its purpose. 
This method is also applicable to other substrates such as alumina, and can be also extended to a multilayered structure by alternately repeating the Zn salt and complex layers.
\subsection*{Optical properties of the crystalline bulk and functionalized TOF}
\begin{figure}[h!]
    \centering
    \flushleft a)\\
    \includegraphics[width=8cm]{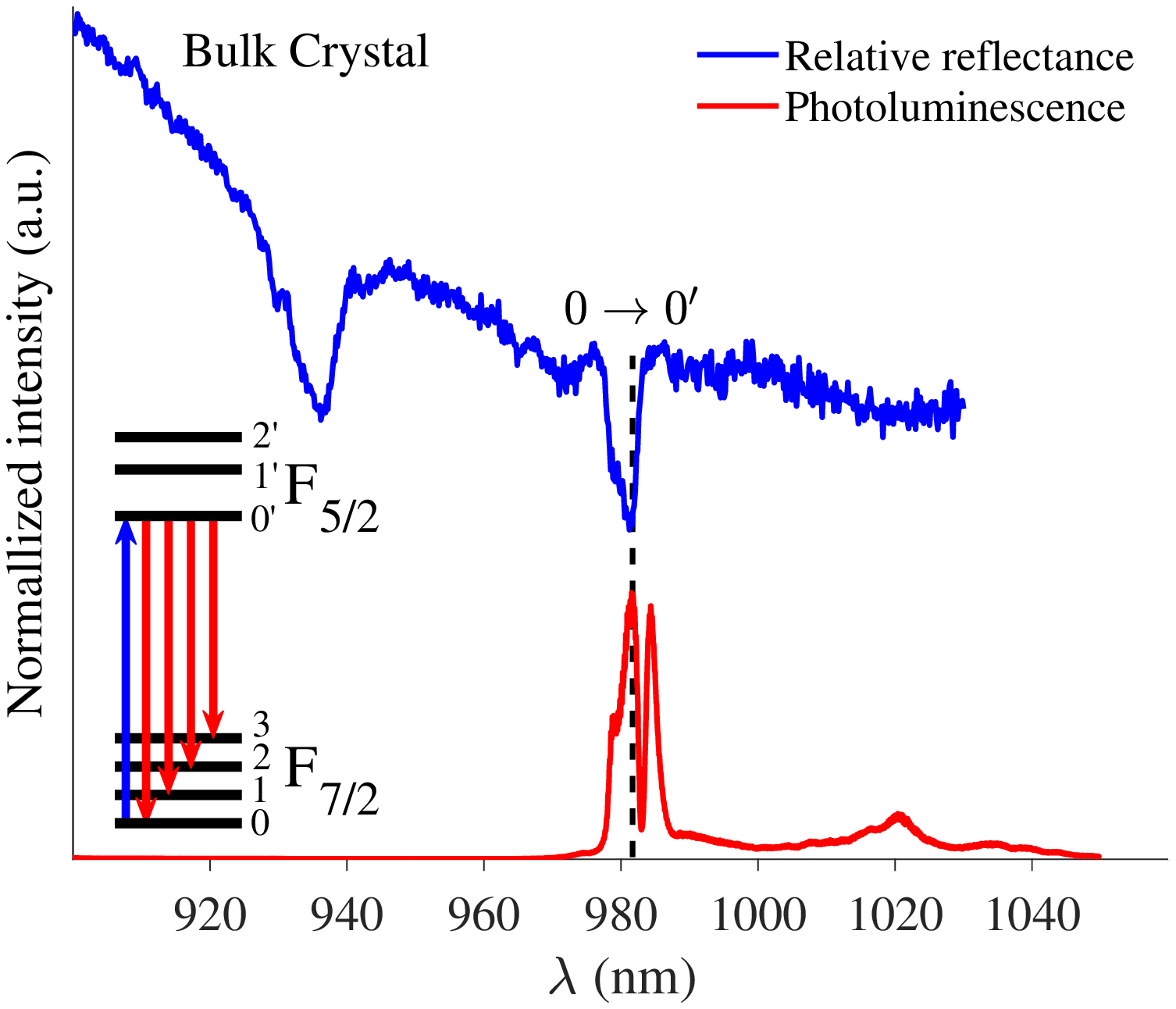}
    \flushleft b)\\
    \includegraphics[width=8cm]{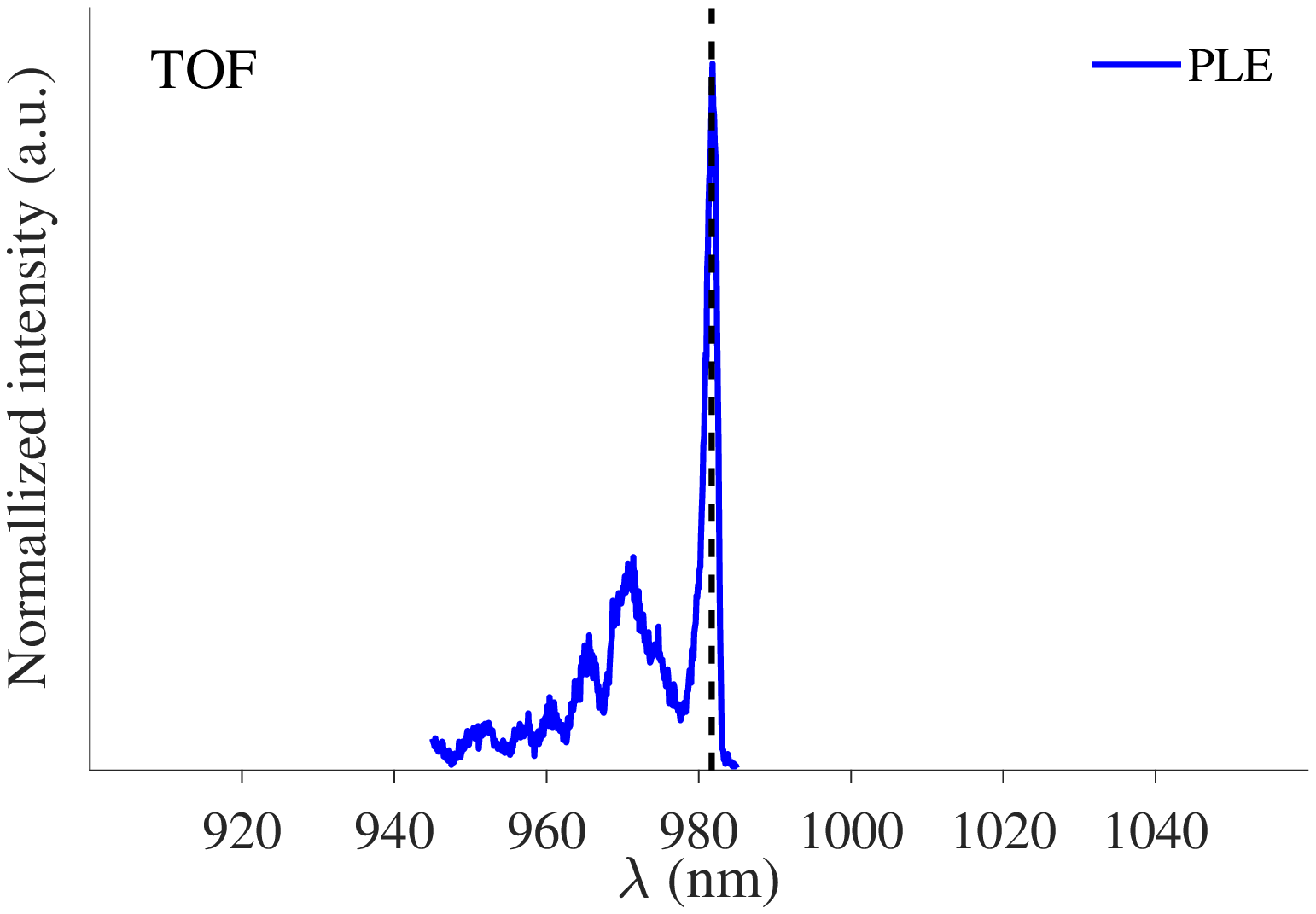}
    \caption{\textbf{Optical spectroscopy of Yb$^{+3}[\mathrm{Zn(II)}_\mathrm{MC}$ (QXA)] complexes at cryogenic temperatures:}(a) Relative reflectance (blue) and PL (red) spectroscopy of crystalline bulk ($\lambda_{ex}=785$~nm) at $T\sim~30~K$, and (b) PLE spectrum of a TOF coated with a monolayer of the complexes at $T < 10~K$. 
    The PLE spectrum was taken within the laser tunability window of 945-985~nm. 
    The main transition width is estimated from a Gaussian model with $\sigma=0.85~nm$. 
    Vertical line at 981.7 nm marks the mutual transition. 
    Therefore, the Yb transitions exhibit negligible Stokes shift, and are insensitive to the deposition method on its own.}
    \label{ref_pl_ple}
\end{figure}

Our primary goal in this work is to demonstrate that the beneficial optical properties of RE complexes can be utilized on a functionalized TOF. 
Therefore, we first explore the optical properties (i.e. relative reflectance and PL - See section D in the SI for the experimental details) of the  Yb$^{+3}[\mathrm{Zn(II)}_\mathrm{MC}$ (QXA)] crystalline bulk  (Fig.~\ref{ref_pl_ple}a) and then compare it to the optical properties of the functionalized TOF (Fig.~\ref{ref_pl_ple}b).

Interaction with the Zero-Phonon-Lines (ZPLs) is crucial for quantum applications and needed for a more precise assessment of the absorption and PL spectra in the crystalline bulk as well as in the monolayer. 
Therefore, we conducted our measurements at cryogenic temperatures.

The PLE experiment was conducted at a temperature below 10 K as indicated by the disappearance of the transition at 984 nm.
The reflectance and PL spectra were obtained using a closed chamber under He atmosphere.
From previous measurements, we estimate the temperature of the chamber at about 30 K (See the SI for details).

We focus on the main transition which consists of inhomogeneously broadened ensemble of Yb ions having spin degenerate pairs (a 4-level system), as well as a hyperfine structure for some of the isotopes. This system is sufficient for most quantum applications.
The blue trace in Fig.~\ref{ref_pl_ple}a is the relative reflectance spectrum of the crystalline bulk. 

The bulk thickness was of a few mm resulting in interference pattern on the spectrum.
The spectrum shows 2 main dips in reflectance at 981.7~nm with a shoulder at 979~nm, and an additional dip at 935~nm. 
Based on the electronic configuration of Yb$^{+3}$ (inset in Fig.~\ref{ref_pl_ple}a) and a previous study on a closely related complex \cite{trivedi2014highly}, we assign the transition at 981.7 nm to the $0\rightarrow 0'$ transition and the other dips to transitions to higher levels in the excited states manifold. 
Additional support to the assignment of the $0\rightarrow 0'$ transition comes from the PL spectrum (red trace in Fig.~\ref{ref_pl_ple}a) that exhibits 2 main emissive transitions at 981.7 and 984.3~nm.
The former transition is very close in frequency to the main transition that is observed in relative reflectance spectrum (marked by a dashed line) indicating that the emission Stokes shift is very small. 
The 984.3~nm transition in the PL spectrum may represent the $0' \rightarrow 1$ transition. 
Weaker peaks are observed at $\sim 1020$~ nm which correspond to transitions to the higher levels in the ground state manifold  $0'\rightarrow 2$ and $0'\rightarrow 3$. 
Next, we characterized the optical transitions in the functionalized TOF. 
To do so, the TOF was connected to a pulsed Photoluminescence Excitation (PLE) spectroscopy optical setup by fiber fusion (splicing)- see Fig.~\ref{device}~a.
The integrated PL signal was collected by Single-Photon Counting Modules (SPCMs) in the backward direction (with respect to the pump; marked by light red arrows), and any parasitic reflection of the excitation pulse was gated using an acousto-optic modulator (AOM) in a single-pass configuration (which means the first $\sim 1 \mu s$ of the backwards signal was blocked, see section E in the SI for more details).

Given the dimensions of the TOF (a $\sim$ 5mm long waist with diameter of $\sim$ 500~nm), the estimated number of emitters is less than $\sim10^{10}$ (see section D in the SI for more details). Note that the geometry and frailty of the TOF prevented us from performing the same measurements we performed on the bulk crystal or on the silicon chip monitor.\\
\begin{figure*}[ht]
    \centering
    \includegraphics[width=8cm]{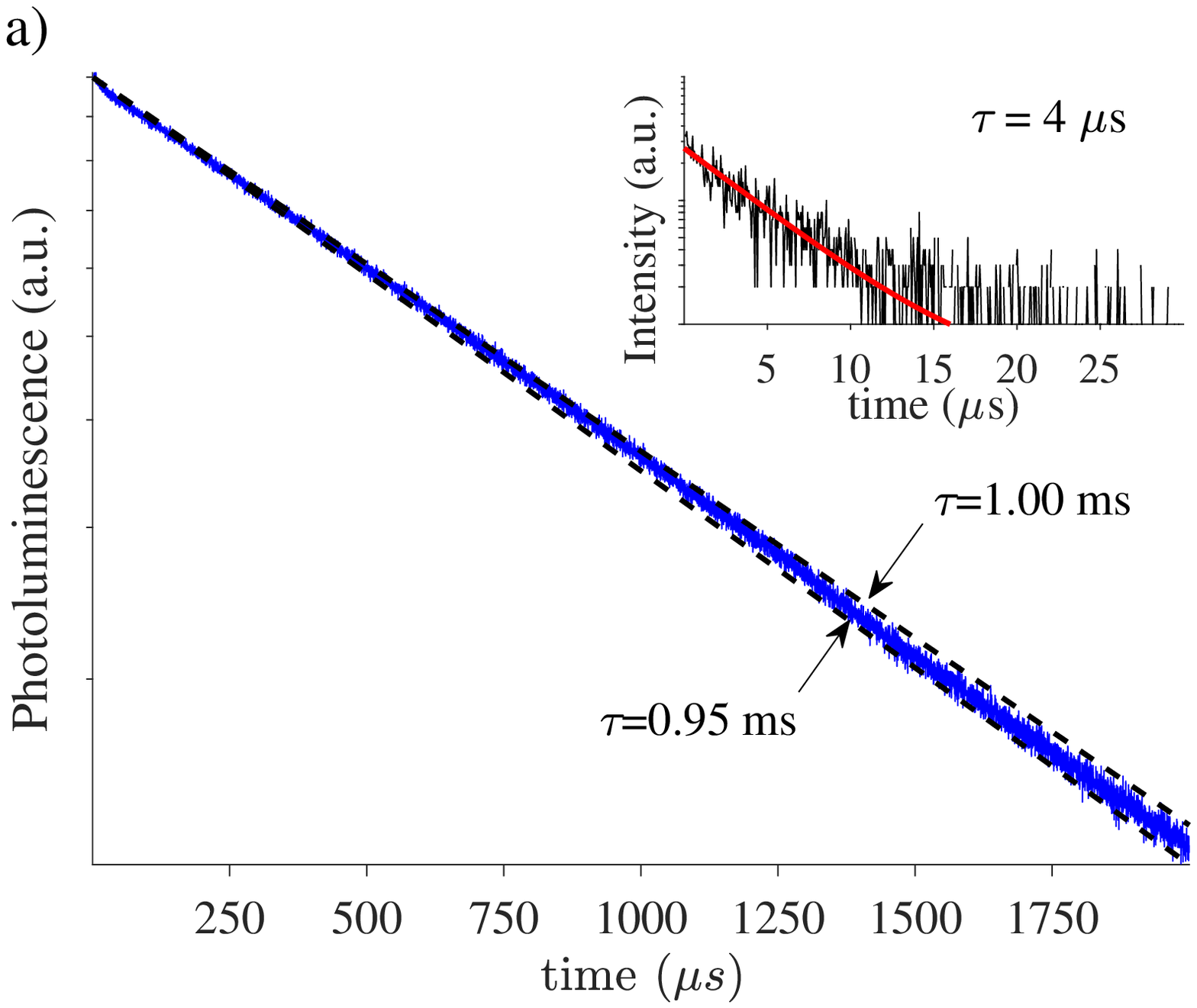}
    \includegraphics[width=8cm]{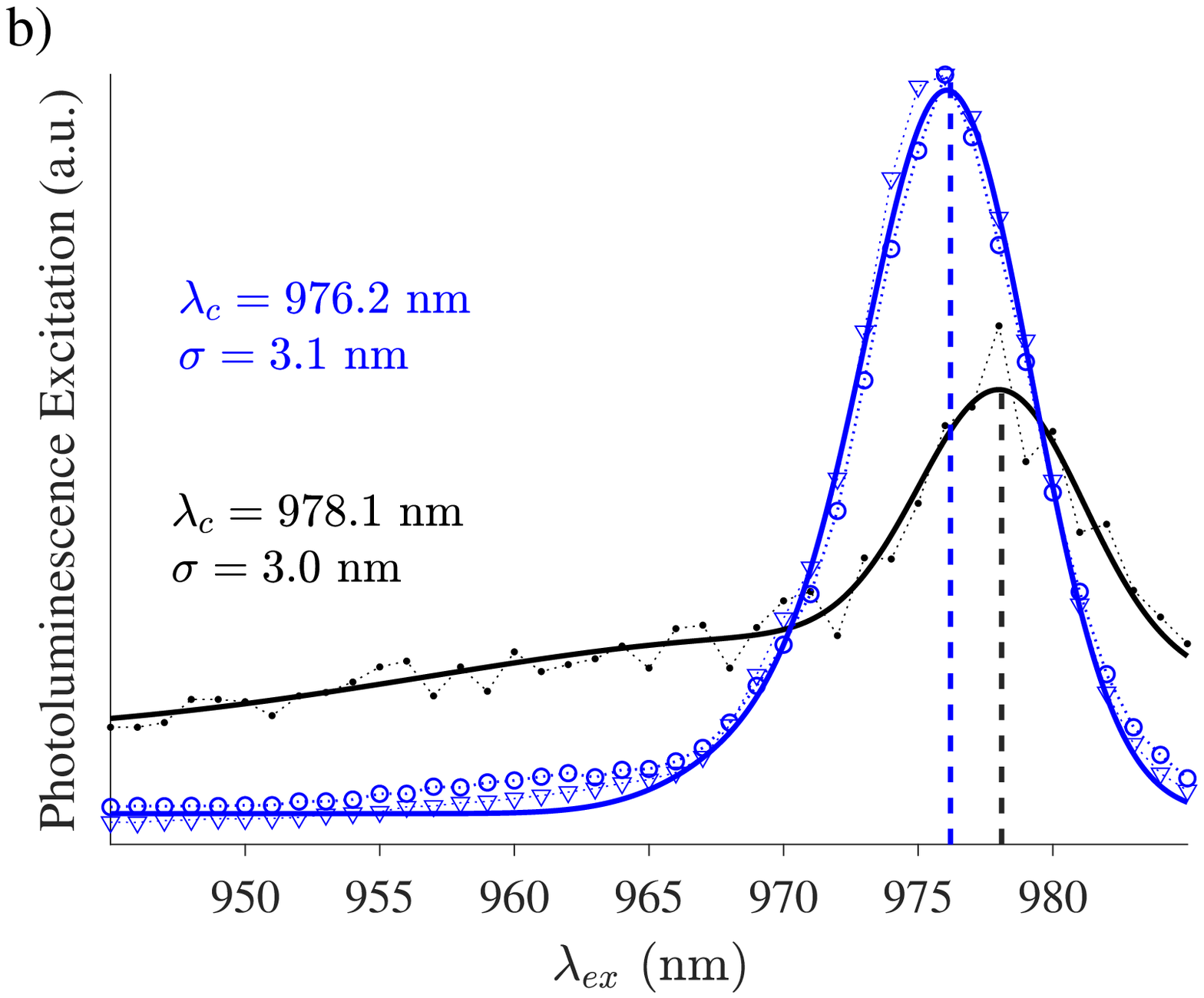}
    \caption{(a) The decay of the PL signal upon excitation on resonance of a TOF functionalized with a monolayer of Yb$^{+3}[\mathrm{Zn(II)}_\mathrm{MC}$ (QXA)] complexes.  
    While at ambient pressure (inset) the monolayer exhibited time constant of a few $\mu s$, after a few days in high vacuum, the time constant exceeds 0.95 ms (blue). (b) PL excitation spectra measurements for different samples of fibers coated with a monolayer of Yb$^{+3}$[Zn(II)$_\mathrm{MC}$ (QXA)]. Solid lines represent fits.
    One of the samples was measured at atmospheric pressure (black dots). Both samples were measured after over a week in vacuum (blue circles and triangles). 
    The black curve was multiplied by a 100 for better visibility. 
    The fits (see SI for details) show a main resonance at 978 nm and 976 nm (at atmospheric pressure and in vacuum respectively). 
    The main peak at 978 nm corresponds to the $0\rightarrow 0'$ transition shifts to 976 nm in vacuum. 
    This is attributed to the removal of ligating solvent molecules which changes the level splitting of the Yb ion.}
    \label{spec_decay}
\end{figure*}

The blue trace in Fig.~\ref{ref_pl_ple}b presents the PLE spectra of the functionalized TOF at cryogenic temperatures obtained by the immersion in a liquid He dewar. 
The spectrum shows a main peak at 981.7~nm. 
A broader feature appears between 960 and 977~nm. 
The most important finding in the context of this study is that the peak at 981.7~nm, which is assigned to the $0\rightarrow 0'$ and marked by a dashed line, is sharp and dominant.
This demonstrated that the absorption spectrum of the Yb$^{+3}[\mathrm{Zn(II)}_\mathrm{MC}$ (QXA)] complex was unaffected by the deposition on the TOF. 
Moreover, it is narrower than the corresponding transition in the crystalline bulk.
We therefore conclude that the two directions (absorption and emission) of the main transition, are effectively unchanged for the two deposition methods. It also possesses a negligible Stokes shift between absorption and emission which implies a relatively pure transition (i.e. uncoupled to vibrations).
The precise assignment of all peaks requires further study of the optical behavior of these complexes.
\par Our aim is to reach a comparable lifetime to Yb inorganic doped crystals, which is on the order of 1~ms (depending on the host crystal) \cite{jacquemet2005efficient,welinski2016high}.

At ambient conditions, the time-resolved PL of the coated fiber exhibits lifetime of 4~$ \mu s$, much shorter than the lifetime of Yb doped inorganic crystals.
Based on the work of Pecoraro et al \cite{trivedi2014highly}, we hypothesized that the origin of the shorter lifetime in this system is the adsorbed solvent molecules on the layer. 
These solvents can be removed by placing the monolayer in vacuum. Since the complexes consist of labile N-O bonds, they are susceptible to thermal decomposition. We therefore avoided bake out at elevated temperatures.

Indeed, by introducing the TOF into high vacuum (as low as 10$^{-7}$ mbar), we extended the lifetime drastically.
The blue traces in Fig.\ref{spec_decay} show the time-resolved PL (a.) and PLE spectrum (b.) of a functionalized TOF after a few days (typically a week) in high vacuum (10$^{-7}$~mbar).
We observe a dramatic increase in lifetime by over two orders of magnitude: from approximately 4 $\mu s$ at ambient conditions to over 0.95 ms. 
Additionally, a distinct shift of the main transition from 978 nm in ambient atmosphere to 976 nm in vacuum indicates a change of the levels splitting resulting from a modified Yb environment.\\

\par
This work presents low-temperature spectra with inhomogeneous linewidth estimated at $\sim 530$ GHz ($\sigma=0.85$ nm fitted with a Gaussian function). This value is 2-3 orders of magnitude larger than the linewidths obtained with low concentration Yb doped crystals, commonly used for quantum applications(e.g. 2.2 GHz in YSO \cite{welinski2016high}
, 0.275 GHz in YVO$_4$ \cite{kindem2018characterization}
, and 3.6 GHz in YAG \cite{bottger2016optical}). 
On the other hand, there are Yb complex such as Yb$(dpa)_3$\cite{reinhard2002high} which show comparable inhomogeneous linewidth in their crystal form at 15 K. 
We believe the wider linewidth in our case is mainly a result of near-field interactions due to the high density of Yb ions in the monolayer (distance of approximately 1.5 nm between adjacent ions, according to the crystal structure).
In terms of lifetime, inorganic doped crystals exhibit lifetime ranging from 0.267 ms \cite{kindem2018characterization}, 0.87 ms \cite{welinski2016high}, to 1.0 ms \cite{bottger2016optical}, and a few Yb doped crystals (such as YLF \cite{puschel2021temperature} and CaF$_2$\cite{camy2007comparative}) exceed the 1~ms lifetime. In organic Yb-based NIR fluorophores, the longest reported lifetime in  is 0.7 ms (via sensitization) \cite{hu2017highly}.
In comparison, our device exhibits 0.95 ms lifetime, which is well within the lifetime range of doped crystals.
Moreover, note that the fluorescence emitted to the fiber is expected to be enhanced due to the fiber's cooperativity (estimated by 0.1-0.2 \cite{le2008cooperative,sague2007cold}). Accordingly, the obtained value of $\sim 0.95$~ms corresponds to an even slightly longer decay time in free space, which indicates that our device exhibits efficient isolation of the Yb$^{+3}$ ions from non-radiative decay channels.\\
 
Interestingly, the monolayer structure also provides a route to control the distance between adjacent RE ions, as well as from the glass surface, thus reducing the effect of decoherence sources originating from the glass.
This platform is generally applicable to all RE ions and can be applied on any photonic device made of oxygenated substrates such as Si, glass or Al\textsubscript{2}O\textsubscript{3}.
That way, it enables enhanced coherent coupling to RE ions utilizing the evanescent field of waveguides and whispering-gallery-mode resonators, potentially opening the path towards a large number of optical and quantum-optical applications.\\
\section*{Conclusions}
In conclusion, we present a new interface between RE ions and TOFs, based on organic complexes with low non-radiative loss, i.e. high intrinsic quantum yield. 
We observed similar spectra of the encapsulated Yb$^{+3}$ intra 4f-transitions in different systems: a crystalline bulk and a monolayer. 
We further observed optical coupling of a monolayer of complexes on a TOF placed in high vacuum, with $\tau > 0.95$~ms, which confirms the efficient suppression of non-radiative decay channels, and is comparable to Yb-doped inorganic crystals.\par
\par
\section*{Available supporting information}
The synthetic procedure for the ligand and complex, as well as the surface functionalization procedure are available.
The X-ray crystallographic coordinates for the structure reported in this study have been deposited at the Cambridge Crystallographic Data Centre (CCDC), under deposition number 2102073. 
The data can be obtained free of charge from The Cambridge Crystallographic Data Centre via\\ \href{https://www.ccdc.cam.ac.uk/data_request/cif}{https://www.ccdc.cam.ac.uk/data\textunderscore request/cif}.
\section*{Acknowledgments}
The authors acknowledge funding from H2020, Excellent Science (DAALI,  899275), IMOD (OR RISHON), Israel Science Foundation, Minerva Foundation, and a
research grant from Dr. Saul Unter.\\
B. Dayan is the Dan Lebas and Roth Sonnewend Professorial Chair
of Physics.\\
O.E.M. thanks Milko van der Boom and his group for the use of their glovebox and some fruitful discussions, Irit Goldian for conducting the AFM measurements, and Oren Tal and Daniel Petukhin for assisting with the measurements in liquid He.
\bibliography{refs}
\end{document}